\documentclass{elsart}
\usepackage{float,latexsym,epsfig}
\usepackage{amsfonts}
\newcommand{\beq}{\begin{equation}}
\newcommand{\eeq}{\end{equation}}
\newcommand{\beqa}{\begin{eqnarray}}
\newcommand{\eeqa}{\end{eqnarray}}
\newcommand{\nn}{\nonumber \\}
\newcommand {\tr}[1]{ \; {\mathop{\mathrm{tr}}\limits_{\quad #1} } \; }
\def \a {\underline{\alpha}}

\def \L {\underline{\Lambda}}
\def \La {\underline{\Lambda}^{(\alpha)}}
\def \Lb {\underline{\Lambda}^{(\beta)}}
\def \s {\sigma}
\def \r {\rho}
\def \e {{\underline{\mathrm{e}}}}
\def \Q {{\underline{\mathrm{Q}}}}
\def \q {\underline{\mathrm{q} }}
\def \t {\tau}
\def \ex {\mathrm{e}}

\def \z {\zeta}
\def \W {W_{1+\infty}\ }

\def \k {\kappa}
\def \D {\Delta}

\def \H {{\mathcal H}}

\def \C {{\mathbb C}}
\def \Z {{\mathbb Z}}

\def \G   {\Gamma}

\def \mod {\; \mathrm{mod}\; }
\def \PF { \mathrm{PF} }
\def\U1{{\widehat{u(1)}}}

\def \ch {\mathrm{ch}}
\def \TO {\mathrm{TO}}

\def \Im {\mathrm{Im} \ }
\def \Re {\mathrm{Re} \ }
\def \df {\stackrel{\mathrm{def}}{=}}
\begin{document}
\begin{frontmatter}
\title{\bf Stability and Activation Gaps of the
Parafermionic Hall States in the Second Landau Level }
\author{L. S. Georgiev\thanksref{INRNE}}
\ead{lg@thp.uni-koeln.de}
\thanks[INRNE]{On leave of absence from Institute for Nuclear Research and 
Nuclear Energy, Tsarigradsko Chaussee 72, BG-1784 Sofia, BULGARIA}
\address{Institut f\"ur Theoretische Physik, 
Universit\"at zu K\"oln,  
Z\"ulpicher Str. 77,  50937  K\"oln, GERMANY}
\begin{abstract}  
Analyzing the effective conformal field theory for the 
parafermionic Hall states,  
corresponding to filling fractions
$\nu_k=2+k/(kM+2)$, $k=2,3,\ldots$, $M$ odd,    
we show that the even $k$ 
plateaux are expected to be more 
stable than their odd $k$ neighbors. 
The reason is that the parafermion chiral algebra can be locally 
extended for $k$ even.
This reconciles the theoretical 
implication, that the bigger the $k$ the less stable the fluid,  
with the experimental fact that,  for $M=1$, the $k=2$ and $k=4$ 
plateaux are already observed at electron temperature 
$T_e\simeq 8$ mK, while the 
Hall resistance for $k=3$ is not precisely quantized  at 
that temperature  in the sample of Pan et al. 
Using a heuristic gap ansatz  we estimate the activation energy gap 
for $\nu_3=13/5$  to be approximately $0.015$ K, 
which implies that the quantization of the Hall conductance could 
be observed for temperature below $1$ mK in the same sample.
We also find an appealing exact relation between the fractional 
electric charge and fractional statistics of the quasiholes. 
Finally, we argue that besides the Moore--Read phase for the 
$\nu_2=5/2$ state there is another relevant phase,  
in which the fundamental quasiholes obey abelian statistics and 
carry half-integer electric charge.
\end{abstract}  
\end{frontmatter}   
\section*{Introduction}  
In 1998 Read and Rezayi  have found \cite{rr} 
a new hierarchy of incompressible ground states, 
in the second Landau level, for 
quantum Hall (QH) fluids corresponding to filling fractions
\begin{equation}
\label{frac}
\nu_k= 2 + \frac{k}{kM+2}, \qquad
\begin{array}{l} k=2,3,4,\ldots \\ M \ \mathrm{odd ,} \end{array}
\end{equation}
(the number $2$ in Eq. (\ref{frac}) comes from the two completely 
filled lowest Landau levels with electrons of both spins).
For what follows, it will be convenient to introduce also the 
particle--hole (PH) conjugated QH fluids 
corresponding to filling fractions 
\begin{equation}
  \label{ph-frac}
  \nu'_k= 2 + \left(1- \frac{k}{kM+2} \right) ,
\end{equation}
where the PH conjugation concerns only the last occupied 
Landau level (assumed polarized).
We shall concentrate on the case $M=1$,  though the general $M$ 
will be kept in most of the formulae.
The first plateau ($k=2$) in the series (\ref{frac}) and (\ref{ph-frac}) 
is the mysterious \cite{willett,eisen88} $\nu=5/2$,  whose nature is still 
unclear and whose ground state
 wave function is believed to have a good overlap with that of the 
 Pfaffian  state \cite{mr,milo-read,morf,cgt,read_5-2}.
The ground state  wave functions for the QH fluids with  filling 
factors (\ref{frac}) have been explicitly 
constructed in \cite{rr} by using the operator product expansion (OPE) of
$\Z_k$ parafermionic currents \cite{zf-para,fat-zam,fat-lyk} describing 
the low-energy effective
field theory  for the edge excitations. A very important feature
 of these wave functions is their cluster decomposition property
\cite{rr,gura,cgt2000pr,cgt2000}, implying that the electrons form 
clusters of $k$-particles  which then interact almost like in a 
Laughlin fluid.

Shortly after the work \cite{rr} Pan et al. \cite{pan} have performed a 
very precise
measurement of the activation energy gap for  $\nu_2=5/2$, at very low
electron temperature, in an extremely high-mobility sample and, for the 
first time, have found that this plateau is quantized to within 2 ppm.
They have also measured the energy gaps corresponding to the strong minima
of $R_{xx}\simeq 0$  at  $\nu_4=8/3$ and  $\nu'_4=7/3$.
In addition, well developed minima at  $\nu_3=13/5$ and
$\nu_5=19/7$ (with PH conjugates  $\nu'_3=12/5$ and
$\nu'_5=16/7$, respectively)
have been seen, however, gap measurements
for these plateaux have not been reported since $R_{xx} \neq 0$. 
Thus, the first plateaux ($k=2,3,4$)
in the sequences (\ref{frac}) and (\ref{ph-frac}) for $M=1$ have been 
experimentally observed \cite{pan}.
A striking fact is that according to the experiment \cite{pan}
 the $k=2$ and  $k=4$ plateaux are clearly quantized while 
the $k=3$  one is not.
On the other hand, the theoretical investigation of the 
stability of the quantum Hall liquids  \cite{fro2000}
(see the  stability criteria S1 and S3 in Sect. 3.3 there)
implies for the hierarchies 
(\ref{frac}) and (\ref{ph-frac}) that 
\textit{the bigger  the $k$ the less stable the fluid}, 
i.e.,
one should expect that the $k=3$ state is more stable
(respectively, has a bigger activation gap)
 than the
$k=4$ one and the former should be ``more visible" than the latter 
at the same temperature.
Here we speak about stability with respect to compression and decompression
of the electron fluid, i.e., with respect to deviation of the magnetic 
field from the center of the plateau. In other words, stability could be 
expressed in terms of the  energy gap of the QH fluid.

In this paper we are trying to  give an explanation
of the above contradiction between theory and experiment. 
More precisely,  using  the conformal field theory (CFT) \cite{CFT-book}
 description of the edge states \cite{wen,fro-ker}, 
 we will show that
\textit{ the even $k$ plateaux are more stable than their odd $k$
neighbors}, i.e., the energy gaps for the former are expected to be 
bigger  than those for the latter.
To this end,  we  make use of recent studies
\cite{cgt2000pr,cgt2000} of the parafermionic Hall states in terms 
of a CFT coset
 construction for the  corresponding edge states and the stability
criteria in \cite{fro2000}. In particular,
we use the chiral partition functions (or CFT characters
for all possible boundary conditions which characterize the universal 
properties of the edge excitations), derived 
in \cite{cgt2000pr,cgt2000}, to 
identify the quasiholes of
the corresponding fluids and find their CFT dimensions, charges, 
fluxes
and fusion rules (or rules for making composite particles).

We have to make  some terminology remarks here. 
We use the CFT as a low-energy (or large-scale)
effective field theory describing the spectrum
(charge, spin and statistics) of the edge 
excitations \cite{fro,wen,fro-ker,fro2000}.
There is a selected class of edge excitations 
which are characterized by the fact that their many-body wave 
functions, which are realized as CFT correlators,   
are single-valued in the (complex plane) coordinates
of all particles. 
In the algebraic approach these excitations are represented by 
{\it relatively local} \cite{fro,fro-ker,fro2000} quantum fields
which contain only integer negative powers of the 
distance in their singular short-distance OPEs  and  
generically form an operator algebra (see, e.g., the condition C4 in 
\cite{fro2000}). 
This algebra completely characterizes the universality class of the
QH fluid and is
called the {\it chiral (observable) algebra} \cite{fro2000}, 
because the excitations on the 
edge are chiral;  all the excitations (charged and neutral) in it 
can be generated from the vacuum by applying only
local chiral  operators. As a consequence, all charged excitations
in that algebra carry integer electric charge (in units $\mathrm{e}=1$) 
and satisfy the standard charge--statistics relation
\cite{fro,fro-ker} for Bose/Fermi particles in the QH effect context; 
in particular, the  relative locality 
with themselves imply that  the CFT dimensions of these fields must be
integer/half-integer. 
The chiral algebra always contains the Virasoro stress energy 
tensor \cite{CFT-book} whose commutation relations are characterized by
a (positive) number $c$,  called the
\textit{central charge}, which is related to 
the Casimir energy on the cylinder \cite{CFT-book}.  
On the other hand, there are
excitations, such as the Laughlin quasihole, whose wave functions are
single-valued in the coordinates of the electrons (or of the chiral 
observables in general) but are multi-valued in their own 
positions\footnote{Usually the Laughlin quasihole wave functions are 
written 
without the factors containing the relative coordinates of the 
quasiholes. In the CFT approach these pieces appear automatically when 
computing the conformal blocks corresponding to the wave functions.}. 
Such excitations cannot be obtained directly from the vacuum 
(by acting with local operators)
but are instead generated by non-local ``{\em intertwiners}" 
\cite{CFT-book}.
The latter fields (called also non-local primary fields) are 
relatively local with the chiral algebra but are
not local with themselves; 
as a consequence, these fields can have 
fractional charges and obey fractional statistics \cite{wilczek}. 
From the algebraic point of view,  such 
intertwiners form {\it representations} \cite{fro2000} of the 
chiral algebra and 
those of them  which have electric charge less than $1$ (in absolute value)
 give the inequivalent 
{\it irreducible representations} (IR)s \cite{fro-ker,fro2000}.
The number of the inequivalent IRs was called the 
\textit{topological order} \cite{wen-top}.
Finally, the CFT describing the edge excitations of the QH fluid is 
expected to be well-defined on the torus and is therefore supposed to be 
$SL(2,\C)$ covariant \cite{fro2000}, i.e., it should be a rational CFT
\footnote{there is a selected class of theories, known as 
$\W$ minimal models \cite{w-min},  which are not rational CFTs
but may  succesfully describe the incompressible QH fluids in the 
principle series}.  
In other words, there is a one--to--one correspondence between the
CFTs and QH universality classes.
Thus, it is always crucial to have a complete list of IRs of the QH 
chiral algebra since they characterize the topological properties of 
all possible quasiparticles in the QH fluid. 

Next, a few words about stability,  justifying  the use of  CFT cosets
for the QH effect,  are in order.
As a first step, one would like to reduce (or gauge away) 
as much as possible  
 the neutral chiral subalgebra
(respectively  the neutral quasiholes' state space) for a QH 
edge since this generically
decreases the Virasoro central charge 
(without changing the QH filling factor)
and, according to the stability criterion S1 in \cite{fro2000},
makes the QH fluid more 
stable. 
In our case \cite{cgt2000pr,cgt2000}, the coset central charge 
\begin{equation}
  \label{c_PF}
  c^{\PF}_k= \frac{2(k-1)}{k+2}
\end{equation}
is smaller than the (neutral part of the) original one
$c^{(0)}_k=2(k-1)$.
On the other hand, however, reducing the algebra in principle 
increases   the number of IRs and, therefore, 
 decreases the fluid's stability (S3 criterion 
in \cite{fro2000}). 
However, after any reduction of the chiral algebra,  some of the
new IRs may  become equivalent,  hence the number of 
inequivalent  ones actually decreases. 
This is exactly what happens in  the coset -- 
due to the very ``symmetric" reduction,   
there are a lot of equivalences, known as 
{\it field identifications under the action of the  simple currents} 
\cite{schw} (see also Appendix B in \cite{fro2000}), which effectively 
reduce the number of IRs and therefore 
the coset fluid  \cite{fro2000,cgt2000} appears to be much more 
stable than the original 
multi-component Luttinger liquid \cite{fro2000}.
This first step for the parafermion QH fluids has been accomplished 
in \cite{cgt2000}.
As a second step towards a bigger stability, one looks for
the maximal (local) extension of the chiral algebra at fixed central 
charge, since this naturally decreases the number of non-equivalent 
IRs and further increase the stability  of the QH fluid, 
according to the stability criterion S3 in \cite{fro2000}.
However, only bosonic and fermionic fields can be used for 
such an extension because of the locality requirement \cite{fro2000}.

Our main result is based on the fact, already observed in \cite{rr},
that for $k=2\kappa$, there are composite fermions/bosons
(for $\kappa$ odd/even) among the $\Z_k$ parafermions 
which are relatively local and therefore can be added to the chiral 
algebra. As a
consequence, this changes the quasihole identification and increases
their CFT dimensions.
We find that the quasihole's CFT dimensions are 
(for $M=1$)
\beq\label{dim}
\D^{\mathrm{q.h.}}_k= \left\{\begin{array}{cl}
\frac{1}{k+2} & \mathrm{for} \ k \ \mathrm{even}   \cr\cr
\frac{1}{2(k+2)} & \mathrm{for} \ k  \ \mathrm{odd}
\end{array}\right. ,
\eeq
i.e., the quasiholes for the even plateaux have twice bigger  
dimensions. A similar relation is found as well for the 
electric charges of the quasiholes. On the other hand, 
Park and Jain \cite{jain} have found that the energy gaps for the 
principle 
series of QH plateaux have universal components proportional to the 
quasiparticle's electric charge. 
{\it We conjecture that the energy gap for the parafermion QH fluid 
has a universal component too}, which should be expressed in terms of 
the universal characteristics of the quasiparticles, 
such as the electric charge  and the quantum statistics;
we find that, for the elementary quasiholes in the parafermion QH 
states the charge is proportional to their CFT dimension (or statistics)  
(see the discussion in Sect.~\ref{sec:gaps}). Therefore, 
we could use Eq. (\ref{dim}) to compute the activation energy
gaps. Taking into account the  gap reduction $\G$ and the 
 renormalization $\alpha$ due to the non-zero thickness, 
Landau level mixing and residual
disorder \cite{jain} we use the following 
{\it measurable gap} ansatz (cf. Eq. (3) in  \cite{jain})
\beq\label{gap}
\widetilde{\D}_k=\alpha \ \D^{\mathrm{q.h.}}_k  \
\frac{e^2}{4 \pi \varepsilon l_k} -\G ,
\eeq
where $\D^{\mathrm{q.h.}}_k$ is given by Eq. (\ref{dim}),
 $l_k = \sqrt{\hbar/eB_k}$ is the magnetic length, $\varepsilon$
is the dielectric constant ($\varepsilon\simeq 12.6 \  \varepsilon_0$ 
for GaAs) 
and $B_k$ is the magnetic field at the
center of the plateau corresponding to the filling factor $\nu_k$
(see Table \ref{tab.1}).
Note that Eq. (\ref{gap}) is not a result of exact or numerical 
diagonalization  of a particular electron interaction. It is based on
the conjecture that the gap has a universal component 
(like in the principle series \cite{jain} where this was found by 
numerical calculation) and on the stability criteria in \cite{fro2000}. 

Fitting the gaps data from \cite{pan}  for the two most stable 
plateaux in the hierarchy (\ref{frac}),  $\widetilde{\D}_2=0.11$ K 
 and  $\widetilde{\D}_4=0.055$ K,  
we find that $\alpha=0.0063$ and $\G=0.045$ K for the sample of \cite{pan}.
This allows us to estimate the energy gaps for the plateaux in the 
hierarchies (\ref{frac}) and  (\ref{ph-frac}) (see Tables~\ref{tab.1} 
and \ref{tab.2} respectively).   
The gap for the $\nu=13/5$ QH state is estimated to be $\sim 15$ mK,
which explains the absence of a Hall plateau in the experiment
\cite{pan}. This also suggests that the quantization is to be observed
for $T_{\mathrm{e}}<1$ mK. 
Next,  
we expect that the less expressed $R_{xx}$ minimum at magnetic field 
$B\simeq 3.35$ T in the same sample   
corresponds to the plateau $11/4$ ($k=6$), whose estimated activation gap
is  $0.029$ K (plus some interference from 
the $14/5$ ($k=8$) state),  rather than to $19/7$ ($k=5$)
as supposed by Pan et al. The plateaux with odd $k> 5$ are not 
expected to be observed in this sample even at zero temperature
and the last observable  even $k$  plateau in this series
is expected to be $14/5$.

In addition, a very interesting property of the parafermion QH fluid
is found: the quasiparticle excitations satisfy a generalized 
charge--statistics relation which explicitly connects the 
fractional electric charge to the fractional quantum statistics.
In fact, we believe that this relation is a fundamental characteristics of
QH quasiparticles.

Finally, according to our analysis of the stability of the parafermion QH 
states, we conclude that, at higher temperature, the $k=2$ state 
may prefer quasiparticles with
fractional charge $1/2$, which obey abelian statistics with $\theta=\pi/2$.
This defines a new universality class relevant for the $\nu=5/2$ QH state,
which is different from that of the Moore--Read (MR) state 
and a (finite temperature) transition between the two phases is possible
(see the discussion in Sect.~\ref{sec:discuss}).

The rest of this paper is organized as follows: in Sect.~\ref{sec:coset}
we review basic facts about the realization of parafermions as a 
coset CFT, in Sect.~\ref{sec:comp} we discuss the consequences of the 
extension of the chiral algebra in terms of the composite fermions/bosons,
in Sect.~\ref{sec:rel} we investigate the relation between charge 
and statistics of QH quasiparticles and in Sect.~\ref{sec:gaps}
we compute the energy gaps for the parafermionic states
using the gap ansatz (\ref{gap}). 
The numerical results are summarized in Tables~\ref{tab.1} and \ref{tab.2}.
%
\section{ The parafermions as a coset }
\label{sec:coset}
The parafermions are realized in \cite{cgt2000pr,cgt2000} (see also  the 
references therein)    as a ``diagonal" 
affine coset \cite{gko} made in an {\it abelian} CFT, which describes
the edge of a multi-component chiral Luttinger liquid.
More precisely, the topological  charge lattice 
(whose metrics is also known as the 
$K$-matrix) is of the type 
$\left( M+2 | {}^1 A_{k-1} {}^1 A_{k-1} \right)$ in the notation
of \cite{fro}, where it was called a {\it maximally symmetric (chiral) 
quantum Hall lattice}. The charge vector $\Q$, whose square
gives the filling factor \cite{fro}, in this $K$-matrix 
model has components $\Q=(1,0,\ldots,0)$ in the {\it dual} lattice 
\cite{fro,cgt2000pr,cgt2000}
implying that this liquid 
possesses a large number of neutral degrees of freedom, which could be  
described by the Kac--Moody algebra \cite{fro,cgt2000pr,cgt2000} 
$\widehat{su(k)_1}\oplus \widehat{su(k)_1}$. 
The $\Z_k$-parafermions \cite{fat-lyk}
are obtained here by the affine coset \cite{gko,cgt2000}
\begin{equation}
  \label{coset}
  \PF_k=
\frac{\widehat{su(k)_1}\oplus \widehat{su(k)_1}}{\widehat{su(k)_2}} ,
\end{equation}
where the ``diagonal" subalgebra $\widehat{su(k)_2}$ in the 
denominator is generated by $J^a=J^a_{(1)}+J^a_{(2)}$ and 
$J^a_{(i)}(z)$ are the currents in the two copies of $\widehat{su(k)_1}$
in the numerator \cite{cgt2000pr,cgt2000}. 
The central charge for the coset (\ref{coset}) 
is given by \cite{cgt2000pr,cgt2000}  Eq. (\ref{c_PF}). 
The above mentioned 
 $su(k)\times su(k)$ symmetry becomes more transparent if we write
the dual pair isomorphism \cite{CFT-book}
\[
\U1 \oplus \widehat{su(k)_1}\oplus \widehat{su(k)_1} \simeq
\widehat{su(2k)_1} \simeq
\widehat{su(2)_k}\oplus \widehat{su(k)_2}.
\]
The $\widehat{su(2)_k}$ Kac--Moody algebra here is simply the usual 
spin  algebra of $k$ pairs of (chiral, complex) Dirac--Weyl 
fermions  corresponding 
to $k$ different layers (one pair for each layer), 
while the $\widehat{su(k)_2}$ is just their layer symmetry. 
Gauging away the latter subalgebra  actually removes the 
layer symmetry, i.e., projects all particles to  the same layer,
which gives another viewpoint for the understanding of the 
clustering  mechanism and the non-abelian statistics of quasiparticles
in these fluids \cite{cgt2000}.

We have to stress that the consistency of the neutral
projection in Eq. (\ref{coset}) requires decoupling of 
neutral and charged degrees of freedom\cite{cgt2000}. However, in the 
original $K$-matrix theory they are not 
completely separated \cite{cgt2000pr,cgt2000}. For the solution of this 
problem the  spin--charge separation is obtained in a {\it decomposable}
sub-lattice of the original one \cite{cgt2000pr,cgt2000}. The price one has to pay
is the appearance of  a
{$\Z_k$ \it pairing rule } (PR) \cite{cgt2000pr,cgt2000} for combining  
neutral and 
charged excitations of the decomposable lattice.
More precisely,  the IRs of the original
$K$-matrix chiral algebra  can be labeled \cite{cgt2000}  
by the (fractional part of the) $\U1$ charge $l$ and the 
$\widehat{su(k)_1}\oplus \widehat{su(k)_1}$ weights $(\L_\mu, \L_\r)$,
where $\L_\mu$, $\mu=0,\ldots,k-1$ are the su($k$) fundamental weights
($\L_0=0$),  such that the PR holds, i.e., $\mu+\r=l \mod k$.
Furthermore, the coset reduction in the neutral part is 
generically  supposed to 
preserve this  PR
 --  now it couples the parafermionic  
$\Z_k$-charge to the electric charge \cite{cgt2000}. 
This inheritance of excitation pairing is one of the main advantages of 
the coset realization (\ref{coset}) of the $\Z_k$ parafermions as 
compared 
to the  more standard $\widehat{su(2)_k}/ \U1$ coset \cite{gep-qiu}. 
Another 
advantage of the diagonal coset approach \cite{cgt2000} is that 
the original model is abelian, i.e., better 
understood, in which  some more complicated non-abelian objects are 
projected out.

The IRs for the coset (\ref{coset}),  which describe the neutral 
excitations of the parafermionic QH fluid,  are labeled by 
sums of two $su(k)$ fundamental weights \cite{cgt2000pr,cgt2000} 
\[
  \L_\mu + \L_\r, \qquad 0\leq \mu \leq \r \leq k-1
\]
(these actually represent all admissible $\widehat{su(k)_2}$ 
weights \cite{CFT-book}).
The conformal dimensions of the corresponding coset  primary fields 
$\Phi^{\PF}(\L_\mu+\L_\r)$
are given by \cite{cgt2000}
\beq\label{CFT-dim}
\D^\PF(\L_\mu+ \L_\rho)=\frac{\mu(k-\rho)}{k}+
\frac{(\rho-\mu)(k-(\rho-\mu ))}{2k(k+2)},
\quad 0\leq\mu\leq\r\leq k-1.
\eeq
As mentioned before, the parafermion fluid 
inherits from the parent 
$K$-matrix model  a $\Z_k$ PR, 
which  can now be  written as follows \cite{cgt2000pr,cgt2000}
\begin{equation}
  \label{PR}
  P=l \mod k ,\quad \mathrm{where} \quad 
P=P\left( \L_\mu + \L_\r \right) = \mu +\r \ \mod k
\end{equation}
and $l$ is the electric charge label introduced above.
This approach allows us to compute all characteristics of the 
parafermionic fluid using those of the Luttinger liquid plus
the coset  reduction.
In particular, the chiral (grand canonical) partition functions 
for the complete parafermionic theory ($l$ and $\rho$ label 
the electric and neutral sectors respectively \cite{cgt2000})
\[
\chi_{l,\r}(\t,\z)= \tr{\H_{l,\r}} \ q^{L_0-c/24} \ 
\ex^{2\pi i \z  J_0^{\mathrm{el}} },
\]
where $L_0-c/24$ is the chiral Hamiltonian and 
$J_0^{\mathrm{el}}$
is the zero mode of the electric current $J^{\mathrm{el}}(z)$
(normalized as usual by 
$J^{\mathrm{el}}(z)J^{\mathrm{el}}(w)\sim \nu_k /(z-w)^2$)
could be written as follows \cite{cgt2000pr,cgt2000}
\begin{equation}
  \label{chars}
\chi_{l,\r}(\t,\z)=\sum_{s=0}^{k-1} K_{l+s(kM+2)}(\t,k\z;k(kM+2))
\ \ch(\L_{l-\r+s} +\L_{s+\r}),
\end{equation} 
where $l$ and $\r$ are respectively $\mod (kM+2)$ and $\mod k$ indices.
Here $K_{\lambda}(\t,k\z;k(kM+2))$ are the $\U1$ (rational torus of 
compactification radius $\sqrt{k(kM+2)}$) chiral partition functions 
(here $q=\ex^{2\pi i \t}$, 
$2\pi R\ \Im \t =1/k_B T$, $2\pi\ \Re \z =V_0/k_B T$, 
$2\pi\ \Im \z= \mu/k_B T$, where $T$ is the temperature,
$k_B$ -- the Boltzmann constant,  
$\mu$ -- the chemical potential and $V_0$ is the Hall 
difference between the edges \cite{cz}),
and $\ch(\L_{\mu} +\L_{\r})$ are the parafermion (neutral) partition 
functions written in the {\it universal chiral partition function} 
(UCPF) form \cite{cgt2000pr,cgt2000,ucpf}
$\ch(\L_{\mu} +\L_{\r})\equiv \ch^{\r}_{\r-\mu}(\t;\PF_k)$,
where $0\leq\mu\leq\r\leq k-1$. The concrete forms of these partition 
functions can be found in \cite{cgt2000pr,cgt2000}, however,
they are not important for our discussion
and we only mention that all the sectors could be obtained uniquely
\cite{cgt2000}  if $0\leq \rho-\mu \leq \rho \leq k-1$.

Each term in the sum in Eq. (\ref{chars}) satisfies the PR (\ref{PR})
and represents some (basic) quasiparticle excitation
\begin{equation}\label{excit}
\Psi_{l,\rho,s}(z)=:\ex^{i \frac{l+s(kM+2)}{\sqrt{k(kM+2)}} \phi(z)  }: 
\Phi^\PF(\L_{l-\rho+s} + \L_{\rho+s})(z)
\end{equation}
of the 
parafermionic QH  fluid.
The electric charge $Q$ and the total conformal 
dimension $\D$ of (\ref{excit}) read \cite{cgt2000}
\begin{eqnarray}
    \label{q.n.a}
    Q&=&k \frac{l+s(kM+2)}{k(kM+2)}= s + \frac{l}{kM+2}, \\
\D&=& \frac{(l+s(kM+2))^2}{2k(kM+2)} + 
\D^{\PF}\left(\L_{l- \r+s} +\L_{s+ \r}\right)  , \label{q.n.b}  \quad
  \end{eqnarray}
where $\D^{\PF}$ is given by Eq. (\ref{CFT-dim}) and $l$ and $\rho$ take the same values as in Eq. (\ref{chars}).
The physical hole operator ($l=\rho=0$, $s=1$) is represented by the tensor product satisfying the PR (\ref{PR})
\cite{cgt2000} 
\[
  \Psi_{\mathrm{hole}}(z)=
  :\ex^{i \frac{kM+2}{\sqrt{k(k M+2)}}\phi(z)}: \ \Psi_1(z)
\]
of the parafermion current $\Psi_1=\Phi^{\PF}(\L_1+\L_1)$ with
the $\U1$ part vertex exponent ($\phi(z)$ being the chiral boson
\cite{CFT-book}
such that $i\partial \phi(z) =J^{\mathrm{el}}(z)/ \sqrt{\nu_k} $ ).
It is characterized by the $s=1$  term in $\chi_{0,0}$ and has 
the following charge, flux (in units of flux quanta) 
and conformal dimension
\begin{equation}
  \label{hole-q.n.}
  Q_{\mathrm{hole}}=1, \quad 
  \Phi_{\mathrm{hole}}= \frac{kM+2}{k}, \quad
\D_{\mathrm{hole}} = \frac{M+2}{2}.
\end{equation}
Notice that the dimension of the hole (respectively, of the electron) is 
independent of $k$ and  for $M=1$ is 
always $3/2$ so that the stability criterion (S2) in \cite{fro2000}
gives no restriction here. 

\noindent
{\bf Remark 1.} 
{\it The fact that the flux (\ref{hole-q.n.}) attached to holes 
(respectively, to electrons) is not integer 
in the vacuum super-selection sector, 
implies that these could not 
live alone, because of flux quantization,  but should rather form 
clusters of $k$ objects \cite{rr,gura,cgt2000pr,cgt2000}
(the only exception is the case $k=2$ where the hole's flux is 
integer and the  edge electrons are not paired \cite{milo-read}).  
We stress that such a phenomenon of electron clustering 
is always supposed to take place when the numerator of the filling 
factor is bigger than $1$, e.g., in the principle series of filling 
fractions. }

Similarly,  the quasihole operator ($l=1$, $\rho=s=0$) is identified 
with  \cite{cgt2000}
\begin{equation}
  \label{q.h.}
  \Psi_{\mathrm{q.h.}}(z)=
 \ :\ex^{i\frac{1}{\sqrt{k(kM+2)}}\phi(z)}: \ \s_1(z) , 
\end{equation}
where $\s_1=\Phi^{\PF}(\L_0+\L_1)$ is  the ``spin" field 
corresponding to the lowest charge and dimension IR of the coset.
It is also worth-noting that it generates all the other IRs by multiple 
fusion
with itself. The field (\ref{q.h.}) is characterized by the $s=0$ term 
in $\chi_{1,0}$ so that its charge and total conformal dimension are
\beq \label{q.h.q.n.}
 Q_{\mathrm{q.h.}}=\frac{1}{kM+2}, \quad
 \Phi_{\mathrm{q.h. }}= \frac{1}{k}, \quad
 \D_{\mathrm{q.h.}}^{(M=1)} = \frac{1}{2(k+2)}. 
\eeq
Again, the quasiholes are enforced to form clusters of $k$
because of flux quantization. 

The electric charge of the quasihole
coincides with the minimal possible charge $Q_{\min}=1/(\ell d_H)$ 
in a QH fluid, according to Eqs. (67--69) in
\cite{fro2000}, where $d_H=kM+2$ is the denominator of the filling factor
and $\ell=1$ is the {\it charge parameter} \cite{fro2000,fro,fro-ker}.
To make contact with the notation in \cite{fro2000} we notice that the 
the conformal dimension of the parafermion $\Psi_1$ is $(k-1)/k$ so that 
the order \cite{fro2000}  of this simple current is 
$\mathrm{ord}(\Psi_1)=k$, while 
 $r(\Psi_1)=2$ coincides with its $\Z_k$-parafermion charge.  
%
%
\section{Composite particles and the maximal chiral algebra extension 
for $\lowercase{k}=2 \k$ }
\label{sec:comp}
As already mentioned in the Introduction, the case $k=2\k$, with
$\k$ a positive integer, is more
delicate. The reason is that, according to Eq. (\ref{CFT-dim}), 
the parafermion
primary field labeled by \cite{cgt2000} $\L_\k+\L_\k$ 
(which is, in fact, the parafermion current denoted by $\psi_\k$ in
\cite{zf-para}) in the neutral sector 
\beq\label{neut}
\Psi^{(0)}_\k=\Phi^{\PF}(\L_\k+\L_\k), \qquad  \D^\PF(\L_\k+\L_\k)=\k /2
\eeq
has a  half-integer/integer  (neutral) conformal 
dimension $\D^\PF$ for $\k$ odd/even.
Next, in order to form an allowed quasiparticle excitation,
this parafermion current must be combined with some charged
operator $:\ex^{i \frac{l}{\sqrt{k(kM+2)}} \phi(z)}:$ 
in such a way that the PR (\ref{PR})  holds,
i.e., 
$2\k = l \ \mod k$, which has only one 
solution\footnote{only for $k=2$, $M=1$ the solutions are two:
 $l=-2,0$} , $l=0$, 
in the interval  $-(kM+2)/2 \leq l \leq (kM+2)/2 -1$ 
(i.e., $l \ \mod (kM+2)$ like in Eq. (\ref{chars}) ) for $M=1$ and $k>2$.
Therefore, $l=0$ defines a unique neutral chiral ``observable" 
{\it composite particle} 
\beq\label{comp}
\Psi_\k(z)= 1 \otimes \Psi_\k^{(0)}(z),
\quad \D_\k=\D^{\PF}_\k=\k/2, \quad Q_\k=\Phi_\k=0  , \quad
\eeq
which could be physically interpreted  
\cite{rr} as made from $\k$ particles 
$\Psi_{\mathrm{p}}=: \ex^{-i\frac{1}{\sqrt{\nu}} \phi(z) }: 
\Phi^{\PF}(\L_{k-1}+\L_{k-1})$ and
$(kM+2)/2$ 
flux \footnote{Note that the field representing a quantum of flux 
 does satisfy the PR (\ref{PR})
(i.e., $0=k \ \mod k$) and is 
therefore a valid excitation of the parafermion QH fluid  } 
$:\ex^{i \frac{k}{\sqrt{k(kM+2)}}\phi(z)}:$. 
These neutral composites  have  
half-integer/integer {\it total} conformal dimensions $\D_\k$, 
are always (super)local with themselves  
(in addition to being local with the entire chiral algebra -- 
see the relative locality discussion in the Introduction)
and by convention should belong to 
the chiral (super)algebra rather than to the set of its representations.
In other words, {\it the original chiral algebra must be extended by the
neutral composite} (\ref{comp}) since it is simply a boson 
for $\k$ even and a fermion for $\k$ odd  according to the standard 
spin--statistics relation $2\D_\k=\theta/\pi \ \mod 2$ ($\D$ in
this case gives the conformal spin). 
Note also that the existence of the field (\ref{comp}) does not contradict
the charge--statistics relation (C3 in \cite{fro2000}) since 
this field is neutral. Moreover, such a field must exist in order to 
restore  the standard charge--statistics relation for the cluster
containing $\k$ electrons. The point is that the flux composite containing
$\k+1$ flux corresponds to electric charge $\k$, however, it is always a 
boson  since its conformal dimension is $\k(\k+1)/2$, which is integer. 
Therefore, to create a cluster of $\k$ electrons one 
needs  a neutral particle, such as (\ref{comp}),  which is a 
boson/fermion when $\k$ is even/odd.

The above extension defines a different rational CFT, which satisfies
all necessary conditions \cite{fro2000} for a QH edge effective field 
theory. Therefore, the extended rational CFT identifies a new universality
class for the $\nu=5/2$, which is different from that for the MR state. 

\begin{figure}[h]
\begin{center}
\epsfig{file=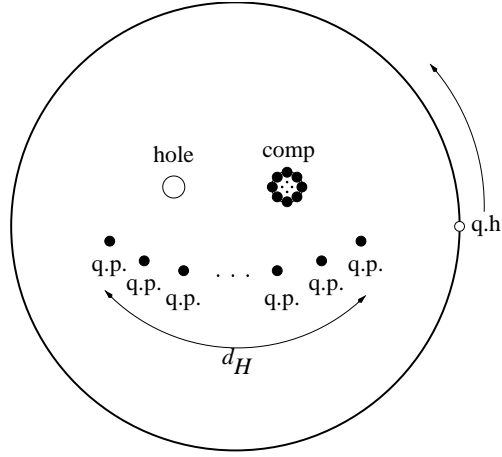,height=6cm,clip=,%
bbllx=0,bblly=0,bburx=460,bbury=412}
\caption{\label{fig:adiab} The cluster consisting of one hole,  
$d_H=kM+2$ quasiparticles  of both types 
(q.p.1 and q.p.2) and one neutral composite (comp)
is topologically equivalent to the pair (1e--1h) and is
completely invisible in the adiabatic transport of  quasiholes (q.h.) 
around it.}
\end{center}
\end{figure}

The neutral composites (\ref{comp}) are fundamental in the sense 
that their correlation
functions are single-valued and, as a consequence of Ward identities, the 
parafermionic component of the stress tensor could be expressed as
\[
T^{\PF}(w)= \frac{k-1}{k(k+2)} \lim\limits_{z \to w}
\left( \frac{\partial}{\partial z} -  \frac{\partial}{\partial w} \right)^2
\left[ (z-w)^\k \Psi_\k(z) \Psi_\k(w)\right].
\]
The importance of the neutral composite (\ref{comp}) stems from the fact 
that the ordinary electron dropped into the QH fluid, for $k=2\k$, splits
into $(kM+2)/2$ elementary quasiparticles of both kind  (q.p.1 and q.p.2
in the lattice description, i.e., before making the coset projection 
\cite{cgt2000pr,cgt2000}) and one composite 
(\ref{comp})
\beq\label{1e}
1 \, \mathrm{e} \simeq \frac{kM+2}{2}(\mathrm{q.p.1}+\mathrm{q.p.2})
+1 \, \mathrm{comp}. 
\eeq
To demonstrate that, we note that the pair (1 hole --- 1 electron)
has no long-range Aharonov--Bohm effects \cite{leinaas} (i.e., it is
characterized by a trivial topological charge) and therefore must be
invisible in the adiabatic transport of any excitation around it.
Consider a cluster consisting of 1 hole, described \cite{cgt2000}  by 
the lattice charge  
$\q_{\mathrm{hole}}=(kM+2)\e^*_1 + \La_1+\Lb_1$, and $(kM+2)/2$ 
quasiparticles of both kinds, described by the lattice charges
\cite{cgt2000} 
$\q_{\mathrm{q.p.1}}= - \e^*_1 + \La_{k-1}$ and  
$\q_{\mathrm{q.p.2}}= - \e^*_1 + \Lb_{k-1}$,
where $\{ \e^*_1,\La_i,\Lb_i , i=1\ldots k-1 \}$ 
is the basis of the dual decomposable 
lattice \cite{cgt2000pr,cgt2000}, with $\La_i$ and $\Lb_i$ being the 
fundamental weights of the two su($k$) algebras,  and  
$\e^*_1=\Q/k$ \cite{cgt2000}.
When we ``fuse" the above mentioned quasiparticles, i.e., when we put 
all  coordinates to the same point, 
the resulting object has a topological charge
\beq\label{excess}
(\k M+1) \left(\q_{\mathrm{q.p.1}}+\q_{\mathrm{q.p.2}} \right)
+\q_{\mathrm{h}}= -\k \left(\La_1+\Lb_1 \right)  ,
\eeq
which is just the sum of the above lattice charges.
The right hand side of Eq. (\ref{excess}) is equivalent to   
$\La_{-\k}+\Lb_{-\k}$  since for the su($k$) 
weights one has\footnote{recall
that two su($k$) weights are equivalent iff their difference belongs 
to the root lattice} 
$a\L_j \simeq \L_{aj \mod k}$,  $a \in \Z$.
Note that the combination with charge (\ref{excess})  carries zero 
electric charge 
(and, respectively, zero magnetic flux), however, 
it is not trivial since $\L_{-\k}$ does not belong to the su($2\k$) root 
lattice. Therefore, a neutral quasiparticle
must be added in order to make the lattice charge (\ref{excess}) trivial;
we note that the neutral composite (\ref{comp}), whose lattice 
charge is $\q_{\mathrm{comp}}=\La_\k+\Lb_\k$, exactly compensates
the topological charge excess in Eq. (\ref{excess}) provided that
\[
\La_{-\k}+\La_{\k}=\sum_{i=1}^{\k-1}i( \a^i+\a^{k-i}) +\k \a^\k 
\simeq 0 
\]
(and similarly for $\Lb$).
In other words,  the topological charge of the cluster 
\begin{equation}\label{cluster}
 1\ \mathrm{comp}+1\ \mathrm{h}+(\k M +1) (\mathrm{q.p.1}+ 
\mathrm{q.p.2})
\end{equation}  
is trivial, i.e.,  
$(\k M+1) \left(\q_{\mathrm{q.p.1}}+\q_{\mathrm{q.p.2}} \right)
+\q_{\mathrm{h}} + \q_{\mathrm{comp}} \simeq  0$, 
so that (\ref{cluster}) is completely invisible in the 
adiabatic transport   of any excitation around it, as shown on  
Fig.~\ref{fig:adiab}; this is equivalent to the statement (\ref{1e}).
After taking the coset this picture does not change since the projection 
is supposed to preserve \cite{cgt2000}  the neutral component 
$\La_1+\Lb_1$ of the 
physical hole  and consequently preserves its 
$\k$-th fusion power.
The difference is now that the quasiparticles of both types become 
indistinguishable and their statistics become 
non-abelian \cite{cgt2000}. Nevertheless, the above adiabatic transport
could be analyzed  in a similar way taking into account
the ``non-abelian" fusion rules \cite{cgt2000} leading to the same 
conclusion that the electron in the $k=2\k$ case has a non-trivial 
neutral component given by (\ref{comp}).
 
\noindent
\textbf{ Remark 2.} 
\textit{ The topological charge of the cluster 
(\ref{1e}), for the parafermion coset (\ref{coset}),  
has components in more than one 
super-selection sector due to the non-abelian fusion rules 
\cite{CFT-book,mr}. In that case, the charge of (\ref{1e}) must be 
projected onto the vacuum sector which is characterized by 
periodic/anti-periodic boundary conditions for the 
bosonic/fermionic currents.} 

One consequence of the chiral algebra extension is that 
the ``extended " QH fluid  is expected to be more stable than the 
original one. The point is that any extended algebra  has 
less (or at least the same number) of IRs, as compared to the original 
one,  since  the IRs of the former must be constructed from those 
IRs of the latter
which are in addition relatively local with respect to  
(\ref{comp}).
According to the stability criterion S3  in \cite{fro2000}, 
reducing the number of admissible IRs would further  increase the 
stability of the QH fluid. 
Let us consider this in more detail. 
The operator product expansion of the 
field (\ref{neut}) with
any coset IR's primary field  $\Phi^{\PF}(\L_\mu+\L_\rho)$ is
determined by the fusion rule \cite{cgt2000}
$\left(\L_\k+\L_\k \right) \times \left(\L_\mu+\L_\r \right) = 
\left(\L_{\mu+\k}+\L_{\r+\k} \right)$  
(all indices are taken $\mod k$) 
so that
\[
 \Psi^{(0)}_\k(z) \ \Phi^{\PF}(\L_\mu+\L_\rho)(w)   \sim
(z-w)^{-\frac{\mu+\rho}{2}}  \Phi^{\PF}(\L_{\mu+\k}+\L_{\rho+\k})(w) ,
\]
where the power of $(z-w)$ is determined  by matching
the CFT dimensions  (\ref{CFT-dim}) of both sides.
Therefore, only those IR's primary fields of the original chiral 
algebra are 
relatively local with (\ref{neut}) which satisfy
\beq\label{restr}
P=\mu+\rho=0 \ \mod 2.
\eeq
This ``even $\Z_k$-charge restriction" reduces half of the IRs of 
the original algebra.
It is not difficult to express the chiral partition functions
$\widetilde{\chi}_{l,\r}(\t,\z)$ corresponding to the extended  
algebra in terms of those (\ref{chars})  for the 
original one. 
First of all, the even-charge restriction (\ref{restr}) implies that
the parameter $l$ in  the original characters (\ref{chars}) must  be 
even since $P(\L_{l- \r+s} +\L_{\r + s})=2s+l$  is supposed to be even.  
Second, since the only field in the extended algebra  which does not 
belong to 
the original one is (\ref{neut}),  naturally we  have
\begin{eqnarray}
  \label{chars-ext}
\widetilde{\chi}_{2l,\r}(\t,\z) &\df & 
\chi_{2l,\r}(\t,\z)+ \chi_{2l,\r+\k}(\t,\z)= 
\sum_{s=0}^{\k-1} K_{l+s(\k M+1)}(\t,\k\z;\k(\k M+1)) \times \nn
\ & \times &  
\left[ \ch(\L_{l-\r+s} +\L_{\r +s}) +\ch(\L_{l-\r+s +\k} +\L_{\r+s+\k}) 
\right] ,
\end{eqnarray}
where we have used the identity \cite{cgt2000}
$K_{2l}(\t,2\z;4m)+K_{2l+2m}(\t,2\z;4m)=K_{l}(\t,\z;m)$.

\noindent
{\bf Remark 3.} 
{\it  Only primary fields with half-integer or integer 
CFT dimensions can be used to extend the chiral algebra. 
Extending the algebra with $\Psi^{(0)}_\k$ (for $k=2\k$, 
$\k$ integer) automatically extends the $\U1$ algebra with
$:\exp\left(\pm i\sqrt{\k(\k M+1)} \phi(z) \right):$
because the corresponding  quantum numbers $Q=\pm \k$, 
$\D=\k(\k+1)/2$ are found
in the vacuum character  $\widetilde{\chi}_{0,0}(\t,\z)$ for $s=0$
and the vacuum module is isomorphic to the chiral algebra.
Therefore, the extension with the field (\ref{neut}) is the  
maximal one.} 

Next, we need to determine the possible values of $l$ and $\r$ 
which give all the independent characters (\ref{chars-ext}). 
We find
\begin{eqnarray}
  \label{range-2l}
  \# \{ 2l\} &=& \# \{ 0,2,4,\ldots, (kM+2)-2 \} = \frac{kM+2}{2} , \quad
 \\ \label{range-rho}
 \# \{ \r \} &=&  
 \# \left\{ \ \r \ \vert \ 0 \ \leq \r \leq \k-\r \right\}=
 \left[\frac{\k}{2} \right] +1,
\end{eqnarray}
where $[x/2]$ stands for the integer part of $x/2$.
The range of $2l$ comes from the even-charge restriction 
(\ref{restr}). To understand how the range  of $\r$ is determined 
we first use the following symmetry  of (\ref{chars-ext})
\begin{equation}
  \label{sym}
\widetilde{\chi}_{2l,\r+\k}(\t,\z) =   
\widetilde{\chi}_{2l,\r}(\t,\z) 
\qquad \Longrightarrow \quad 0 \leq \r \leq \k-1
\end{equation}
to reduce the number of non-equivalent choices of $\r$, and 
next consider only the $s=0$ term in (\ref{chars-ext}) 
keeping in 
mind that the other terms are obtained by the action  of the 
simple currents (expressed by the $\mod \ \k$ sum in 
(\ref{chars-ext})) and represent the same IR of the full 
$\U1\times \PF_k$ theory.  
In addition, since we are interested in the neutral 
degeneracy, we can fix an arbitrary  value of charge, for example
 $2l=0$, and  consider  the $s=0$ neutral character in 
$\widetilde{\chi_{0,0}}$
 \begin{equation}
   \label{rho-range}
   \left[\ch\left(\L_\r +\L_{k-\r} \right)+
\ch\left(\L_{\r+\k} +\L_{\k-\r} \right) \right] ,
 \end{equation}
which could be labeled by the smallest of the   4 $\L$'s indices.
We have $\r < k-\r$ and 
$\k-\r \leq \k+\r $,  due to the range (\ref{sym}),  so that the character 
(\ref{rho-range}) could be 
labeled by the smaller of the numbers $\r, \k-\r$. Thus we find the 
restriction for $\r$ to be $0\leq \r \leq \k-\r$ which gives 
(\ref{range-rho}).

One important characteristics  of the universality class
of any QH fluid  is the topological order \cite{wen-top}. 
In the $k=2\k$ case
this number is given by 
\begin{equation}
  \label{TO-even}
  \TO(k=2\k)=\# \{ 2l\} \# \{ \r \} = 
(\k M+1) \left(\left[\frac{\k}{2} \right] +1  \right) ,
\end{equation}
which is just the product of the two ranges (\ref{range-2l}) and  
(\ref{range-rho}). 
We stress that this number is {\it smaller} than the 
general-$k$  number (conjectured in \cite{rr} and derived in 
\cite{cgt2000pr,cgt2000})
\begin{equation}
  \label{TO-gen}
   \TO(k)=\frac{(kM+2)(k+1)}{2} 
\end{equation}
computed for $k$ even, which is the main  justification of the chiral 
algebra 
extension  presented above since, according to the  stability
criterion S3 in \cite{fro2000}, 
{\it the less this number the more stable the fluid}.
(Note that the topological order for $k$ odd is still given by
 Eq. (\ref{TO-gen}).)

Now, we can see the difference between the odd-$k$ and the even-$k$ 
cases.  The 
coset primary field with the lowest non-zero charge and dimension
 for $k$ odd 
 is labeled by $\L_0+\L_1$, 
while for  $k$ even it is $\L_0+\L_2$; the reason is that   
$\L_0+\L_1$ does not satisfy the even-charge restriction 
(\ref{restr}), i.e.,
it is not relatively local with respect to (\ref{neut}).
Therefore, the dimensions of the quasiholes for the $k$-even case
would be given by a formula different from Eq. (\ref{q.h.q.n.}).
Indeed, since the field $\Phi^{\PF}(\L_0+\L_2)$ is characterized by 
the $s=0$
term in $\widetilde{\chi}_{2,0}(\t,\z)$ the elementary charge and 
CFT-dimension is (according to Eq. (\ref{q.h.}) for $M=1$, and 
Eq. (\ref{q.n.a}) and (\ref{q.n.b}) for $s=0$, $l=2$,  $\r=0$)
\beq\label{even-dim}
  Q^{\mathrm{q.h.}}_{k=2\k} = 
\frac{2}{k+2}, \qquad 
\D^{\mathrm{q.h.}}_{k=2\k} = 
\frac{1}{k+2}.
\eeq
For clarity, the quasiholes' charges and dimensions for the first 
10 plateaux in the 
parafermionic hierarchies (\ref{frac}) and (\ref{ph-frac}) 
are given in Table~\ref{tab.1}.

We stress here that the quantum numbers $P$ in Eq.~(\ref{PR}), 
of the parafermionic currents $\Phi^{\PF}(\L_\mu+\L_\mu)$
in the $k=2\k$ case generate only  an
even-charge subgroup $\Z_\k \subset \Z_{2\k}$. 
In this case, however, one can redefine the parafermion charge (\ref{PR})
as follows
\beq\label{P_even}
P'(\L_\mu+\L_\r) \df  \frac{P(\L_\mu+\L_\r)}{2} \ \mod k, 
\eeq
for $\mu+\r=0 \ \mod 2$,  in agreement with Eq.~(\ref{restr}).
The new charge $P'$ belongs to $\Z_{2\k}$ and is preserved under 
the fusion 
rules \cite{cgt2000}. 
In addition, for all IRs satisfying Eq.~(\ref{restr}),  
one can define a \textit{chiral fermion parity} 
according to $\gamma_F=P' \ \mod 2$.

In particular, 
the composite particles (\ref{comp})
have a $\Z_k$ charge $P(\L_\k+\L_\k)=2\k=0 \mod k$, while 
their modified parafermion charge (\ref{P_even}) is 
$P'(\L_\k+\L_\k)=\k$ and  thus $\gamma_F=\k \ \mod 2$,
in agreement with the discussion after Eq.~(\ref{comp}).
%
\section{Relations between charge, spin and  statistics of 
QH quasiparticles}\label{sec:rel}
Despite the fact that the electrons in a QH fluid are non-relativistic 
their low-energy  collective excitations (quasiparticles and quasiholes)
 are believed \cite{wilczek} to satisfy a 
\textit{generalized} spin--statistics relation 
\cite{leinaas} 
\beq\label{s-s}
2\D_{\mathrm{q.h.}}=\frac{\theta_{\mathrm{q.h.}}}{\pi}=
S_{\mathrm{q.h.}} + S_{\mathrm{q.p.}} , 
\eeq 
(in general this must be  a  $\mod \Z$ relation, however, for the 
elementary 
quasiholes (\ref{q.h.}) and quasiparticles it is exact \cite{leinaas}), 
where the CFT 
dimension $\D_{\mathrm{q.h.}}=\D_{\mathrm{q.p.}}$ is actually the average
spin of the pair (quasiparticle--quasihole) and 
$\theta_{\mathrm{q.h.}}$ is the 
statistical angle. Eq.~(\ref{s-s}) explains why the CFT dimensions 
can be identified with the quantum statistics of the  quasiparticles.

On the other hand, there is a very interesting relation between the charge
and the statistics of the QH quasiparticles corresponding to the 
Read--Rezayi states, namely, the electric charges (\ref{q.n.a})
and CFT dimensions (\ref{q.n.b}) are related 
(we set $M=1$ for simplicity) by
a \textit{generalized charge--statistics relation}
\begin{equation}\label{c-s}
2 \D = Q + \frac{2n}{d_H} ,
\end{equation}
where $d_H=k+2$ is the denominator of the filling factor and
\begin{equation}\label{2n}
2n = d_H \left[ \mu(\mu+1) +2 b \mu \right] + 2b(\rho' -\mu) +  kb(b-1)
\end{equation}
is an even integer. 
Given the parameters $l$, $\r$ and $s$ of the excitation (\ref{excit}),
the new  parameters $\mu, \r' \in \Z_k$, and  $a,b,c \in\Z$ entering  
Eq.~(\ref{2n}) are uniquely determined from 
\begin{eqnarray}
\min \left(l-\rho+s, \rho+s \right)&=&\mu+ak, \quad b = s-\mu+a +c \nn
\max \left(l-\rho+s, \rho+s \right)&=&\rho'+ck . \nonumber 
\end{eqnarray}
Note that the elementary quasiholes 
($l=1$, $\rho=0$, $s=0$ for $k$ odd and 
$l=2$, $\rho=0$, $s=0$ for $k$ even) satisfy 
Eq. (\ref{c-s}) with $n=0$, i.e.,
\begin{equation}\label{c-s-q.h.}
2 \D_{\mathrm{q.h.}} = Q_{\mathrm{q.h.}}.
\end{equation}
The elementary quasiparticles
($l=-1$, $\rho=0$, $s=0$ for $k$ odd and 
$l=-2$, $\rho=0$, $s=0$ for $k$ even) satisfy 
Eq. (\ref{c-s}) with $n=1$, and $n=2$ respectively, however,
in both cases one has
\begin{equation}\label{c-s-q.p.}
2 \D_{\mathrm{q.p.}} = \vert Q_{\mathrm{q.p.}} \vert.
\end{equation}
Both Eqs. (\ref{c-s-q.h.}) and (\ref{c-s-q.p.})
together with the spin--statistics relation (\ref{s-s})
explicitly relate the fractional charge of the (elementary) 
QH quasiparticles to their fractional
statistics \cite{wilczek}. 
This looks natural because spin in 2D 
is an additive quantum number 
(like the electric charge), though 
this additivity, which in CFT language  is called 
``matching CFT dimensions",  is masked in OPEs 
by the powers of the distance (in other words, spin can partially 
transform into relative orbital momentum when building a composite 
particle). Nevertheless, for ``indecomposable" 
primary fields, 
such as the quasihole,  which 
cannot be generated in the right hand side of an OPE of other fields 
with smaller CFT dimensions,
this additivity amounts to a linear relation between charge and spin. 
We stress that 
the quantum statistics of a quasiparticle is usually computed by 
Berry phase technique \cite{mr,wilczek}, however, this computation 
explicitly uses the 
quasiparticle wave functions that  are in general not known.
The advantage of the generalized charge--statistics 
relation (\ref{c-s-q.h.})
is that it determines the quasihole's CFT dimension (respectively, 
statistics) in terms of
the minimal electric charge, which is fixed by the filling factor.
This gives another viewpoint on the relation between the precisely
quantized (fractional) Hall conductivity and the sharp 
quantization of the (fractional)  electric charge.

It is important to note that general  multi-electron clusters that are 
bosons/fermions for even/odd electric charge satisfy Eq. (\ref{c-s}) 
with $n= m d_H$ ($m\in\Z$) thus reproducing  the standard 
charge--statistics 
relation (see e.g. Eq. (31) in \cite{fro2000}). 

Finally, we expect that the generalized charge--statistics relations
(\ref{c-s}), (\ref{c-s-q.h.}) and (\ref{c-s-q.p.})
go beyond the scope of the Read--Rezayi states in the second Landau 
level. We believe that such relations are fundamental characteristics of
the QH fluids \cite{5-2} and we expect them to hold also in 
the principle series $\nu=n/(2n+1)$ of QH filling 
fractions.
\section{Energy gaps for the  parafermion QH fluids}
\label{sec:gaps}
In this section we compute the energy gaps for
the parafermionic Hall states in the second Landau level.
To this end, we use Eq. (\ref{gap})  with $\D_{\mathrm{q.h.}}$ derived 
from the effective field theory at the edge. 
Before going into the details of the calculation we have to make 
several remarks. First of all, the number 
$(2n+1)^{-1}$, which appears in 
the gap formula Eq. (3) in \cite{jain}, is generically the charge of the 
quasihole. In our case, however, this charge  turns out to be
twice  the CFT dimension of the 
quasihole operator (see Eqs. (\ref{c-s-q.h.}) and 
(\ref{c-s-q.p.})) which justifies the ansatz (\ref{gap}). 
It is worth-recalling
the ``conventional wisdom" \cite{jain} that (i) the non-zero thickness
in the transversal direction and Landau level mixing simply renormalize
$\alpha$  in (\ref{gap}) (ii) the Landau level broadening due to 
the residual disorder
simply reduces the gap by $\G$ and (iii) $\alpha$ and $\G$ are 
approximately independent of $k$ (resp. of $\nu_k$).
Second, we recall that usually the conformal dimensions are related 
to the energies of the edge excitations,  whose effective chiral 
Hamiltonian is \cite{cz}
$v(L_0-c/24)/R$ , where $L_0$ is the zero mode of the Virasoro 
stress tensor,   $v$ is the edge states velocity and 
$R$ the radius of the edge circle, 
i.e., all excitations become gapless in the 
thermodynamic limit $R\to\infty$. 
However, since this CFT
describes the edge of an \textit{incompressible} QH fluid,
where the bulk charge-density waves are suppressed by an energy gap, 
this  means that any 
localized charged  excitation in the bulk is completely 
transmitted to the edge
and is seen there as an edge excitation carrying  all quantum numbers 
(such as topological charges and  quantum statistics) of its bulk 
counterpart. The point is that the ideal gap is proportional to some 
universal quantity, such as the quasiparticle's electric charge, 
which can be computed at the edge. 

We speculate that  Eq.~(\ref{gap}) admits an interpretation of the 
energy gap as 
the centrifugal barrier for creating the minimal spin dictated by 
the statistics of quasiparticles, which, on the other hand, is 
determined by the filling factor. Indeed, if we assume that the 
quasiparticle energy 
gap is proportional to its spin, then
the average gap $\epsilon$ of the pair (q.p.--q.h.) is proportional to the 
average spin, which, according to the spin--statistics relation
(\ref{s-s}), is exactly equal to the quantum statistics \cite{leinaas},
respectively, to the CFT dimension of the quasihole
\[
\frac{\epsilon_{\mathrm{q.p.}}+ \epsilon_{\mathrm{q.h.}}}{2}\sim
\frac{S_{\mathrm{q.p.}}+ S_{\mathrm{q.h.}}}{2} =  
\frac{\theta_{\mathrm{q.h.}}}{2\pi} = \D_{\mathrm{q.h.}}. 
\]
In realistic cases, one has to take into account the non-universal 
effects (see (i), (ii) and (iii) above) as in Eq.~(\ref{gap}).
Next, the charge--statistics relations (\ref{c-s-q.h.}) and 
(\ref{c-s-q.p.}) relate the 
(ideal) energy gap to the electric charge of the elementary  
quasiparticle, which is equal to the denominator of the filling factor,
thus reproducing Eq.~(3) in \cite{jain}.

Now, let us go back to Eq. (\ref{gap}). The two constants 
$\alpha$ and $\G$  could be determined by fitting the measured 
energy gaps for the two most 
stable plateaux $k=2$, $\widetilde{\D}_2=0.110$ K and
$k=4$, $\widetilde{\D}_4=0.055$ K. We find that
\begin{equation}
  \label{param}
  \alpha=0.0063, \qquad \G= 0.045  \ \mathrm{K} ,
\end{equation}
which allows us to compute the other activation gaps in  the hierarchies
(\ref{frac}) and (\ref{ph-frac}) for the sample of \cite{pan}. 
Note that $\alpha$ and 
$\Gamma$ (depending on the particle density and layer thicknes) 
are not universal and must be re-fitted for any other sample.

The more important data about the plateaux (\ref{frac}) 
in the sample of \cite{pan}
  is summarized in Table~\ref{tab.1}. In  Table~\ref{tab.2} we give the
similar data for the PH conjugate plateaux (\ref{ph-frac}).
In both tables $M=1$, the average electron density 
$n_e\simeq 2.2 \times 10^{11}$ cm$^{-2}$ is taken from
\cite{pan}, the magnetic field, $B_k$ for the hierarchy(\ref{frac})
and  $B'_k$ for the hierarchy(\ref{ph-frac}),  
is measured in T. 
The quasiholes' charges $Q^{\mathrm{q.h.}}_k$ are measured in units
in which the electron charge is $-1$.
The CFT dimensions of the quasiholes are $\D^{\mathrm{q.h.}}_k$,  
$\mathrm{TO}(k)$  is the topological order and  the measurable 
activation energy gaps  
$\widetilde{\D}_k$ and $\widetilde{\D'}_k$ are given in ${}^{\circ}$K. 
The gaps are 
given to 2 significant figures because the gaps for
$k=2$ and $k=4$ were measured with this precision 
and in order to be able to compare these predictions with the experiment. 
The values of the quasiholes' charges $Q^{\mathrm{q.h.}}_k$, 
CFT dimensions $\D^{\mathrm{q.h.}}_k$  
and topological order $\mathrm{TO}(k)$ in Table~\ref{tab.2} are the 
same like those corresponding to the same $k$ in Table~\ref{tab.1}.
%
%
\begin{table}[ht]
\caption{Quasiholes' charges  and CFT dimensions,  
topological orders and  the measurable activation energy gap  
for  the parafermionic states at filling factors (\ref{frac}) for 
the sample of \cite{pan}.
  \label{tab.1} }
\begin{tabular}{c||cccccccccc}
\hline 
$k$ & $1$ & $2$ & $3$ & $4$ & $5$ & $6$ & $7$& $8$ & $9$ & $10$ \\
\hline\hline 
$\nu_k$ & ${7/3}$ & ${5/2}$  & ${13/5}$ 
& ${8/3}$ &  ${19/7}$ & ${11/4}$ & 
${25/9}$ & ${14/5}$ &${31/11}$ & ${17/6}$ \\ 
\hline
$B_k$ & $3.92$ & $3.66$  & $3.52$ & $3.43$ & $3.37$ & $3.32$ & $3.29$ 
& $3.26$ & $3.24$ & $3.23$ \\ 
\hline 
$Q^{\mathrm{q.h.}}_k$  & ${1/3}$ & ${1/2}$ & ${1/5}$
& ${1/3}$ & ${1/7}$ & ${1/4}$ & ${1/9}$
& ${1/5}$ & ${1/11}$ & ${1/6}$ \\
\hline
$\D^{\mathrm{q.h.}}_k$ & ${1/6}$ & ${1/4}$ & ${1/10}$& ${1/6}$
& ${1/14}$ & ${1/8}$ & ${1/18}$& ${1/10}$ & 
${1/22}$ & ${1/12}$ \\
\hline
$\mathrm{TO}(k)$ & $3$  & $2$& $10$ & $6$ & $21$ & $8$ & $36$ & $15$ 
& $55$ & $18$ \\
\hline
$\widetilde{\D}_k$  & $0.062$   & $0.110$ & $0.015$  
& $0.055$  & $-0.003$ 
 & $0.029$  & $-0.012$  & $0.013$  & $-0.019$  & $0.003$ \\ 
\hline
\end{tabular}
\end{table}
%
%
%
\begin{center}
\begin{table}[htb]
\begin{center}
\caption{The measurable activation energy gap 
for  the PH conjugate plateaux of the parafermionic states 
at filling factors (\ref{ph-frac}) for the sample of \cite{pan}.
  \label{tab.2} }
\begin{tabular}{c||cccccccccc}
\hline 
$k$ & $1$ & $2$  & $3$ & $4$ & $5$ & $6$ & $7$& $8$ & $9$ & $10$ \\
\hline\hline 
$\nu'_k$ & ${8/3}$& ${5/2}$ & ${12/5}$ & ${7/3}$ 
&  ${16/7}$ & ${9/4}$ & ${20/9}$ & ${11/5}$ &
${24/11}$ & ${13/6}$ \\ 
\hline
$B'_k$& $3.43$& $3.66$  &  $3.81$ & $3.92$ & $4.00$ & $4.06$ & $4.11$ 
& $4.15$ & $4.19$ & $4.22$ \\ 
\hline 
$\widetilde{\D'}_k$ & $0.055$ & $0.110$ & $0.018$  & $0.062$  & $0.001$ 
 & $0.037$  & $-0.09$  & $0.021$  & $-0.015$  & $0.010$ 
\\ 
\hline
\end{tabular}
\end{center}
\end{table}
\end{center}
The $k=2$ plateau is common for 
both Tables since it is self-conjugate. The $8/3$ plateau appears
as $k=4$ in Table \ref{tab.1} and as $k=1$ in Table \ref{tab.2}, 
while the plateau $7/3$ appears in reverse order.
Notice that the $k=1$ states are actually a Laughlin states, as
mentioned in \cite{rr}. This is directly seen from the $k$-body 
interaction \cite{rr}, as well as from the coset (\ref{coset}),
which gives trivial neutral part and the remaining CFT is  
$\widehat{u(1)}$ with compactification radius $3$.
We see that, in general, the gaps for the PH-conjugate states 
(\ref{ph-frac})
are slightly bigger than those for (\ref{frac}) because the 
magnetic length 
in the former is smaller than that in the latter. However,
the experimental gap ($0.10$ K) for $\nu'_4=7/3$ is much 
bigger \cite{pan} than  the gap ($0.055$ K) of its conjugate, 
the $\nu_4=8/3$ one, which cannot be 
explained by the smaller magnetic length only.
This can be explained by the fact that the Laughlin states for
$k=1$ have bigger gaps than the $k=4$ parafermion states, according to the
stability criterion S3 in \cite{fro2000}, because the
topological order of the former is $3$ while that for the latter is $6$
(after the extension).
Nevertheless, this makes no difference in Eq.~(\ref{gap})
since the electric charges of the quasiholes of both $k=1$ and $k=4$ 
states are equal.

We stress that it is in first approximation  that
$\alpha$ and $\Gamma$  in Eq. (\ref{gap}) are independent of the 
filling factor, so that one cannot expect a perfect overlap 
with the experiment.
If we accept, comparing to the result in  \cite{jain}, that the accuracy 
is not better than $30 \%$ then  the gap for the plateau 
at $\nu=13/5$ should be 
$15\pm 5$ mK, while that for $\nu=7/3$ is $62\pm 20$ mK. Nevertheless,
this approximation seems to be sufficient to determine the  structure 
of plateaux in the parafermionic hierarchy.
Note that without the extension of the chiral algebra for $k$ even,
the gap for $\nu=13/5$, computed by  our gap ansatz (\ref{gap}),
  would be $\simeq 77\pm 25$ mK, which obviously contradicts
the experimental data \cite{pan}, which suggests that this value is 
$\sim 10$ mK.

Taking into account that the electron temperature $T_{e}$ at which
$\nu_2=5/2$ is already quantized is \cite{pan}  $T_{e}\simeq 8$ mK, 
we expect  that the  $\nu_3=13/5$ plateau quantization 
could be observed at $T_{e}\simeq 1$ mK in the same sample.
In addition, the $k=8$ state ($\nu_8=14/5$) should be
approximately  as stable as the $k=3$ one ($\nu_3=13/5$) 
(apart from the difference in the magnetic length) and 
is also expected to be observed at that temperature. 

We also conjecture that the $R_{xx}$ minimum around $B\simeq 3.35$ T
corresponds to  $\nu_6=11/4$ rather than to  $\nu_5=19/7$
and  more precise measurements down to electron temperatures around   
$1.5$ mK would be necessary in order to
confirm this. 
In the same way,  the $R_{xx}$ minimum around $B\simeq 4.05$ T 
corresponds $\nu'_6=9/4$ rather than to $\nu'_5=16/7$.
The plateaux with $k>5$ as well as the $\nu=19/7$ one 
($k=5$ in (\ref{frac})) are not expected to be observed in 
the sample of \cite{pan},
even at zero temperature, since the gap reduction $\G$ in Eq. (\ref{gap}) 
is bigger than the (renormalized) ideal gap. The plateaux with even 
$k=10$ ($\nu=17/6$ and $\nu'=13/6$) are neither expected to be seen, 
despite the fact that their  gaps are big enough,
because of the overlaps with the wide integer Hall plateaux at  
$\nu=3$ and $2$ respectively.
The last plateau in the series (\ref{frac}) observable
in the sample of \cite{pan} is expected to be $14/5$ ($k=8$) 
 for the hierarchy (\ref{frac}) and 
$11/5$ ($k=8$) for (\ref{ph-frac}). 

An important message to the experiment is to measure 
the activation gaps for a high-mobility sample at temperatures  
down to $1$ mK,  which seem to be accessible \cite{ertl}.
We disagree with the conclusion in \cite{pan} that ``the gaps for 
$\nu=5/2$, $8/3$ and $7/3$ are approximately equal and on the order of 
$2$ K". 
Therefore,  it would be interesting to check experimentally  our 
predictions for the gaps at 
$\nu=7/3$, $13/5$, $12/5$, $16/7$, $11/4$, $9/4$, $14/5$ and $11/5$.
%
\section{Summary and discussion}
\label{sec:discuss}
We have demonstrated that the non-monotonic plateaux structure of 
the parafermionic QH states in the second Landau level 
can be understood in terms of the 
effective CFT for the edge states.
The main idea is to find the maximal chiral algebra extension since
 this increases
the stability (respectively, the energy gap) of the QH fluid,
according to the stability criterion S3 in \cite{fro2000}.
However, only 
bosons and fermions   can be used for that purpose and the key observation 
is that for even $k$ there are neutral composite bosons/fermions 
among the excitations of the parafermion QH states, 
which can extend the algebra.
This reduces the topological order and increases the minimal electric 
charge, respectively, the activation energy gap.

The detailed analysis of the spectrum of the parafermion edge excitations
shows that, in addition to the generalized spin--statistics relation
(\ref{s-s}), 
there is an exact charge--statistics relation for the elementary 
quasiparticles (\ref{c-s-q.p.})  and quasiholes (\ref{c-s-q.h.}). 
These two relations together with the 
assumption that the pure gap is proportional to the minimal electric 
charge, like in the principle series of filling factor \cite{jain}, 
allow us to compute 
the measurable activation gaps for all plateaux in the hierarchies 
(\ref{frac}) and (\ref{ph-frac}) for the sample of \cite{pan}.
In simple words, assuming that the minimal electric charge for
$k$ even is twice bigger, gives an explanation of the non-monotonic
structure of the parafermion plateaux.

We stress that the gap reduction $\G$ in Eq. (\ref{param}), 
that we computed for  
the sample of \cite{pan}, appears to 
be  2 orders of magnitude
less than the value estimated in \cite{pan}.
  Although it seems plausible that this non-universal quantity is smaller
for the extremely high quality sample of \cite{pan}
(since it depends more on the non-zero thickness than on the residual 
  disorder \cite{jain})
we have to make clear that if the realistic gap reduction is bigger, 
then our argument becomes qualitative. In other words, 
even if Eq. (\ref{gap})
is not a good approximation, the fact that the extended fluids
for $k$ even are more stable than  the original ones, still may 
explain the 
non-monotonic structure of the observed plateaux.
In order to check our predictions new activation experiments, 
down to electron temperature of $1.5$ mK, are necessary.

For the $\nu=5/2$ state ($k=2$) our stability analysis implies that 
the elementary quasiholes have electric charge $1/2$ and obey 
\textit{abelian} fractional statistics  $\theta=\pi/2$.
Then the ordinary electron splits into two quasiparticles plus one 
Majorana fermion at the edge \footnote{This is not surprising since
even in the $K$-matrix theory the quasiparticle topological charge
is not collinear with that of the electron (see Eq.~(\ref{1e})) so that a 
neutral component is always present.}.
 The ground state wave function is
still given by the Pfaffian \cite{mr,milo-read,read_5-2,cgt}, however, 
the structure of the quasihole excitations is different \cite{5-2}. 
Thus, the extended rational CFT discussed in Sect.~\ref{sec:comp} 
defines a new universality class for the $\nu=5/2$ QH state, which is
clearly distinguished from the MR one.
On the other hand, recent numerical calculations \cite{5-2hr} have shown 
that the numerical ground state for the pseudopotential Hamiltonian has an
excellent overlap with the MR state, which means that most likely 
at $T=0$ the true $\nu=5/2$ QH state belongs to the MR universality class.
However, one thing characterizing the MR state is that 
the $\Z_2$ symmetry corresponding to
 the  conservation of the chiral fermion parity is spontaneously broken
\cite{5-2}. As temperature increases, this symmetry may be restored
driving the system into the abelian phase  described above
(in which chiral fermion parity is conserved). Thus, one may expect a 
phase transition \cite{5-2} between the two phases, which could  
be easily detected since the minimal electric charge changes 
from $1/4$ to $1/2$. This may possibly explain the kink at $15$ mK  
observed in the activation experiment \cite{pan}. 
 
Thus, we conclude that precise quasiparticle charge-measuring 
experiments, such as shot-noise, especially for temperature in the range
$10\div 30$ mK and a high-mobility sample similar to that of \cite{pan}, 
as well as new activation experiments down to temperature $1$ mK, 
 could be really decisive for revealing the nature of the enigmatic 
$\nu=5/2$ state.
\begin{ack} 
I would like to thank Andrea Cappelli, Ivan Todorov and
J\"urg Fr\"ohlich
for fruitful discussions, 
Wei Pan for kindly offering details from the experiment
and 
the Max-Planck-Institute f\"ur Physik Komplexer Systeme in Dresden
for hospitality and financial support. This work was 
supported by DFG 
through Schwerpunktprogramm ``Quanten-Hall-Systeme"
under the program
``Konforme Feldtheorie der Quanten-Hall-Plateau-\"Uberg\"ange".
\end{ack}
\def\NP{Nucl. Phys. }
\def\PRL{Phys. Rev. Lett.}
\def\PL{Phys. Lett. }
\def\PR{Phys. Rev. }
\def\CMP{Comm. Math. Phys. }
\def\IJMP{Int. J. Mod. Phys. }
\def\JSP{J. Stat. Phys. }
\def\JP{J. Phys. }
\bibliography{gaps}
%
%
%
\end{document}